\documentclass[twocolumn]{aastex63}
\usepackage{multirow}
\usepackage[T1]{fontenc}
\usepackage{mhchem}
\usepackage{url}
\usepackage{bbding}
\usepackage{pifont}
\usepackage{wasysym}
\usepackage{amssymb}

\usepackage{graphics,graphicx}
\usepackage{color, colortbl}

\newcommand{\thCO}{\(^{13}\textrm{CO}\)}
\newcommand{\CetO}{\(\textrm{C}^{18}\textrm{O}\)}

\newcommand{\CsvO}{\(\textrm{C}^{17}\textrm{O}\)}

\newcommand{\thCetO}{\(^{13}\textrm{C}^{18}\textrm{O}\)}

\definecolor{LightYellow}{rgb}{1,.98,.804}
\submitjournal{APJ}

\begin{document}
%\begin{linenumbers}   #this worked
\title{Complex Organics Surrounding the FU Ori-Type Object V1057 Cyg Indicative of Sublimated Ices}

\shorttitle{NOEMA Survey of Outbursting YSOs}
%\shortauthors{}

%anon
%\correspondingauthor{Jenny Calahan}
%\email{jcalahan@umich.edu}

\author[0000-0002-0150-0125]{Jenny K. Calahan}
\affiliation{University of Michigan, 323 West Hall, 1085 South University Avenue, Ann Arbor, MI 48109, USA }
\affiliation{Center for Astrophysics | Harvard \& Smithsonian, 60 Garden St., Cambridge, MA 02138, USA }

\author[0000-0003-4179-6394]{Edwin A. Bergin}
\affiliation{University of Michigan, 323 West Hall, 1085 South University Avenue, Ann Arbor, MI 48109, USA }

\author[0000-0002-2555-9869]{Merel van't Hoff}
\affiliation{University of Michigan, 323 West Hall, 1085 South University Avenue, Ann Arbor, MI 48109, USA }

\author[0000-0002-0150-0125]{Alice Booth}
\affiliation{Center for Astrophysics | Harvard \& Smithsonian, 60 Garden St., Cambridge, MA 02138, USA }

\author[0000-0002-0150-0125]{Karin \"{O}berg}
\affiliation{Center for Astrophysics | Harvard \& Smithsonian, 60 Garden St., Cambridge, MA 02138, USA }

\author[0000-0003-4001-3589]{Ke Zhang}
\affiliation{Department of Astronomy, University of Wisconsin-Madison, 475 N. Charter St., Madison., WI 53706}

\author[0000-0002-3950-5386]{Nuria Calvet}
\affiliation{University of Michigan, 323 West Hall, 1085 South University Avenue, Ann Arbor, MI 48109, USA }

\author[0000-0003-1430-8519]{Lee Hartmann}
\affiliation{University of Michigan, 323 West Hall, 1085 South University Avenue, Ann Arbor, MI 48109, USA }

\begin{abstract}

FU Ori and EX Lup type objects present natural experiments for understanding a critical stage in the star and planet formation process. These objects offer insight into %episodic accretion onto protostars, mass transfer from the envelope to the star, and 
the diversity of molecules available to forming planetary systems due to a sudden increase in accretion and central luminosity causes the disk and surrounding material to increase in temperature. This allows for volatiles to sublimate off of grains and exist in the gas-phase for tens to hundreds of years post initial outburst. While this dynamic stage may be common for solar-type protostars, observations of the chemical impact of these bursts are rare. In this article, we present observations from the NOrthern Extended Millimeter Array (NOEMA) of five Young Stellar Objects (YSOs) that have undergone outbursts within the past 100 years and catalog the volatile chemistry found within $\sim$1000~au of the YSO. Only one source clearly shows a line rich spectra with $>$11 molecules detected including complex organics and water, as is an expected spectra signature for a post-outburst source. This source is V1057 Cyg, and we present it as the northern analog to the well studied and molecule-rich FU Ori source, V883 Ori. Our conclusions on the chemical inventory of the other four sources in our sample are sensitivity limited, as V1057 Cyg contains the highest disk/envelope gas mass. %Additionally, optically thick mm-dust may hinder our view of the molecular emission originating from the compact disk regions of these sources. %Our current observations do not constrain the other four sources as being distinctly different as V1057 Cyg and V883 Ori, as the other disks/envelope masses are lower than V1057 Cyg and dust observation may plague these observations. 

  %In addition to the mass determinations, we identify bright lines captured over NOEMA's wide bandwidth. In all sources we see multiple lines of SO and \ce{H2CO} in addition to the common CO isotopologues. One source, V1057 Cyg shows 48 individual transitions among 12 different molecular species. The disks and central regions around outbursting young stellar objects appear to be massive objects, often containing envelope emission putting these sources on the edge of Class I/II sources.

%\usepackage{}%adding an additional factor of uncertainty when converting \HtwoetO{} abundances to that of \HtwoO{}.  
%The presence of this surface layer with elevated heavy oxygen isotope abundances carried by water vapor in the inner disk provides a reservoir of material that could contribute to puzzling meteoritic enrichment's of \etO{}.
\end{abstract}


\section{Introduction} \label{sec:intro}

The majority of young stellar objects likely undergo episodic accretion events as they evolve onto the main sequence \citep[i.e.]{Hartmann97,Audard14}. Extreme changes in luminosity were captured towards YSOs with infrared excesses as early as the 1970s \citep{Herbig77}. These objects would undergo sudden and intense brightening in observed magnitudes, translating to a rise in 10-100 times their original luminosity. These objects are called FU Ori-type objects (FUOrs) after their namesake source FU Ori. Two other FUOrs identified early-on are V1057 Cyg and V1515 Cyg, and since then the sample of ``classic'' FUOrs has expanded to $\sim$15 \citep{Audard14, Connelley18}. FUOr-like objects do not have an observed increase in luminosity but have FUOr-like features in their spectra \citep[i.e., CO absorption in the mid-IR, `peaky' and rich water vapor lines, absence of emission lines; see][ for a detailed summary]{Connelley18}. After the initial quick rise, a slow decline in luminosity takes place, taking at least 100 years for a source to reach quiescence. Each source has a different morphology in its decline in brightness over time, with some source fading quickly at first, then platauing over time (V1057 Cyg). Others decline in brightness much more slowly (FU Ori) \citep{Hartmann96}.

EXOr outbursts are a different classification of outbursting event towards YSOs and their surrounding disk. These sources also exhibit rises in observed brightness, but the increase in their luminosity is less dramatic (factors of 2--10), and the rise and fall of their bolometric luminosity completes after a much shorter period of time ($\sim$1-10 yrs). All in all, there are 40+ known eruptive young stars \citep[see ][]{Szabo22}, with a recent increase in FUOr/EXLup candidates due to missions like Gaia and other wide-field telescopes \citep[e.g.,][]{Tran24, Giannini24}.  

The increase in luminosity in both FUOrs and EXOrs is attributed to sudden increases in the accretion rate from the disk onto the star, with accretion rates reaching 10$^{-4}$--10$^{-5}$ M$_{\odot}$/yr for FUOrs \citep{Hartmann96, Audard14, Fischer23}. The exact mechanism that triggers FUOr outbursts is not known, however it likely involves one or more of the following: (1) a gravitationally triggered magnetorotational instability \citep{Hartmann96, Liu16, Cieza18} (2) a thermal instability caused by trapped heat in high opacity regions \citep{Bell94, Armitage01,Boley06,Zhu10} (3) fragmentation of the disk in the outer regions and/or periodic infall from a surrounding envelope \citep{Vorobyov06, Vorobyov13, Lee24}(4) an interaction with a near-by stellar companion or massive planet \citep{Bonnell92}. 

Sources identified as FUOrs have been classified as both Class I and II objects, however most likely fall between the Class I-II stage. Many (but not all) FUOrs appear to have a remaining envelope surrounding the disk itself, and are located in active star forming regions. Lying between the Class I-II phase coincides with the advent of planet formation. Looking at disk structure, disks around Class 0/I source exhibit little to no rings, gaps, or asymmetric structures \citep{Ohashi23} while Class II protoplanetary disks nearly always have disk substructure when seen in the continuum \citep{Andrews18}. At least some of these substructures are presumed to be a direct or indirect result of Jupiter-mass protoplanets already coalesced within Class II protoplanetary disks, insinuating that the planet formation process starts early and happens quickly. Thus, these initial stages of planet formation very likely occur along side energetic outburst events as is found in FUOrs. This makes these sources critical data points in gaining an understanding of both star and planet formation. 

An FUOr-type outburst is a natural laboratory for revealing the chemical content of the ice reservoir within planet-forming disks, and for tracking chemical timescales. A convenient consequence of a sudden burst of accretion is that the surrounding disk will undergo a rapid increase in temperature. The volatile chemical reservoir that is typically frozen out onto the surfaces of dust grains will sublimate and become a part of the more-easily-observable gas-phase population of a disk. The volatile content that is locked onto the ice mantel surrounding dust grains within roughly 10 au is the main volatile reservoir for forming planetary cores, and an outburst event offers an extremely rare opportunity to catalog this volatile planet-forming material. Thus, FUOr-type outbursts offer a glimpse into the chemistry of forming planets. V883 Ori is the best example of this, an FUOr source that has proven to be volatile rich with tens of different molecules detected in this source alone \citep{vantHoff2018_V883Ori,Lee19,Lee24,Tobin23,Yamato24}. It is thought that a source like V883 Ori is not uniquely-rich in these volatiles, but the outbursting conditions allow for typical disk chemistry to be exposed. V883 Ori has been thoroughly observed source, and as of recently, is one of the only classic FU Ori source known to be highly line-rich. L1551 IRS 5 is another FUOr-like source that appears to be rich with molecular lines \citep{Marchand24}. HBC494 is yet another FUOr object that has been observed over a wide bandwidth with NOEMA, with some detections of \ce{CH3OH} and \ce{CH3CN}, but otherwise has a line-poor spectrum \citep{vantHoff24}.

These two sources currently are the only two sources observed with high sensitivity, across a wide bandwidth, with interferometers, and these FUOrs are thought to represent `typical' Class I/II sources. They provide data points for COMs, sulfur-rich species, and water; connecting the dots between bright, line-rich protostars, a few Class II protoplanetary disks, and comets. Other FUOr sources should be surveyed so that additional data can fill in this knowledge gap. 

In this paper, we present a survey of five outbursting sources observed with the NOrthern Extended Millimeter Array (NOEMA\footnote{Based on observations carried out with the IRAM Interferometer NOEMA. IRAM is supported by INSU/CNRS (France), MPG (Germany) and IGN (Spain)}). We present the chemical inventory of each source, focusing on the most molecule-rich source in our sample, and present upper limits on the non-detections for the other four sources.

% Additionally, the outburst has the potential to strongly influence the environment of the system, affecting the stellar evolution, disk formation, and subsequent or possibly on-going planet formation \citep{Zhang15,Harsono18}. 
%The environment and properties of FU Ori objects, focusing on the surrounding envelope and the embedded disk, are essential knowledge for a comprehensive understanding of this key stellar evolutionary phase and planet formation. 

% It has been suggested that an FU Ori, or similar type accretion outburst, can explain observed CO snowlines at large radii towards low luminosity objects \citep{Hsieh18}. 

%FU Ori objects also present a unique glimpse into the full chemical inventory available to actively forming planets. Planets form out of the refractory material, volatile ices and gas. Due to the fact that most Class II protoplanetary disks are below 50~K, most of the volatile reservoir is frozen-out onto grains and thus unobservable in radio wavelengths. 
%Given the wide frequency range of NOEMA, we also have the opportunity to identify molecular transitions outside of CO. 

\section{Methods}\label{sec:method}

\begin{deluxetable*}{cccccccc}
\label{tab:SourceProperties}
\tablecolumns{5}
\tablewidth{0pt}
\tablecaption{Source Properties}
\tablehead{Name & Classification & R.A. & Dec & Gaia & Stellar Asso. & Outburst & L$_{\rm bol}$ Peak \\
& & & & Distance [pc] & Distance [pc] & Date & [L$_{\odot}$]
}
\startdata
V1057 Cyg	&	FU Ori	& 20h 58m 53.733s	& +44d 15\arcmin~ 28.389\arcsec & 907$^{+19}_{-20}$ & 795	&	1970	&	250-800	\\
V1735 Cyg	&	FU Ori	& 21h 47m 20.663s & +47d 32\arcmin~ 3.857\arcsec &	691$^{+35}_{-38}$ & 752	&	1957-1965	&	235	\\
V1647 Ori	&	Peculiar & 5h 46m 13.137s & -0d 6\arcmin~ 4.885\arcsec	&	413$^{+20}_{-22}$ & n/a	&	1966, 2003, 2008	&	34-44	\\
V1515 Cyg	&	FU Ori	& 20h 23m 48.016s & +42d 12\arcmin~ 25.781\arcsec&	902$^{+12}_{-13}$ & 960	&	1950	&	200	\\
V2492 Cyg	&	UX Ori/EX Ori	& 20h 51m 26.237s	& +44d 05\arcmin~23.869\arcsec & 804$^{+24}_{-26}$	& n/a &	2009-2010	&	43	\\
\enddata
\tablecomments{Distances taken from parallaxes in Gaia Release 3 \citep{Gaia3}, however it is worth noting that outburst sources are particularly difficult to determine distances to. Other work use the assumed stellar associations to determine distances to these sources, and can disagree with GR3 by 100 parsecs \citep[see][]{Kuhn19}. Outburst date and luminosity taken from review \cite{Audard14} and references therein. Stated ranges indicate the timespan where the onset of the outburst could have started.}
\end{deluxetable*}

\subsection{Source Selection}
The goal of our observational survey was to identify sources that would be ideal candidates to find a chemical signature indicative of an outburst.  The best candidates will be bright and massive sources, thus, we isolate objects that have been published and observed with NOEMA or the Atacama Large Millimeter Array (ALMA) in \CetO{} $J$=1-0 or 2-1, and are located within the observational range of NOEMA \citep{Feher17,Principe18,Abraham18}. There are nine total sources that fit this criteria; we chose the brightest five sources, each with \CetO{} emission peaking directly on source. Within the northern hemisphere, these objects have the highest likelihood of a optically thin mass tracers as well as molecules expected to be sublimated off of dust grains. More source information, and their properties can be found in Table~\ref{tab:SourceProperties}.

\subsection{NOEMA Observations}

These results are derived from NOEMA observations that took place July - October 2021 as a part of project S21AL. Each source was observed for $\sim$3.4 hrs with a beam sizes $\sim$ 1.$^{\prime\prime}$58 $\times$ 1.$^{\prime\prime}$50 (V1735 Cyg, V1057 Cyg) 2.$^{\prime\prime}$45 $\times$ 1.$^{\prime\prime}$69 (V1647 Ori) 1.$^{\prime\prime}$04 $\times$ 0.$^{\prime\prime}$92 (V2492 Cyg, V1515 Cyg). We used Band 3, specifically in the range of 202-211 GHz (lower sideband) and 218-226 GHz (upper sideband). This setup simultaneously covers the \thCO{}, \CetO{}, \CsvO{}, and \thCetO{} J=2-1 transitions at 2 MHz resolution. We placed high velocity windows (62.5 kHz) over a select number of lines including the CO isotopologues, \ce{CH3CN}, \ce{H2^{13}CO}, HDO, \ce{H2CS}, SO, and \ce{SO2}. These lines were expected to be bright, thus line wing morphology can be probed.  Further data reduction details and results from the CO isotopologues are detailed in \citet{Calahan24}. %The initial round of data quality assurance and reduction was done with a staff member at NOEMA. The raw data was self-calibrated and CLEAN-ed using the GILDAS package \textsc{mapping}\footnote{https://www.iram.fr/IRAMFR/GILDAS}. The first step was to find the systematic velocity and correct the UV-tables to the accurate values, which aids in line detection. Next we isolate the continuum by zero-ing out every line and other feature such as dead pixels or end-of-band wiggles. With the continuum left over, we average over every $\sim$200 channels to compress down to a 2D continuum image, that is then self-calibrated and imaged. 
%We then CLEAN that image over a number of iterations that leads to a plateau in the total flux in the image. Once the continuum image is found, the self-calibration can be done. First we find a solution containing a number of iterations and clean components that leads to an increase in the S/N level and a plateau in total flux. 
%That solution is then imposed on the original spectra that contains the emission lines. The self-calibrated data is then continuum-subtracted isolating and imaged. For the goal of determining the mass of each source, we utilized the high-velocity resolution (0.5 km/s or 2 MHz) spectra and of \ce{^{13}CO}, \ce{C^{18}O}, \ce{C^{17}O} and \ce{^{13}C^{18}O}. V1735 Cyg likely contains significant extended emission in the three brightest CO isotopologues, as its solution contains beam artifacts that proved to be less strong if the smallest baselines were excluded. However, to maintain consistency between sources and lines, the full NOEMA configuration including all baseline was used for the final results. %To identify lines, the full lower side band (LSB) and upper side band (USB) were used and CLEAN-ed using the same solution. A mask was used for each CLEAN solution that encircled the central mass and decreased artifacts outside of where emission was known to exist. 

\section{Results}\label{sec:results}

\begin{deluxetable*}{cccccccc}
\label{tab:Common lines}
\tablecolumns{7}
\tablewidth{0pt}
\tabletypesize{\small}
\tablecaption{Identified Lines Across all Outburst Sources}
\tablehead{Molecule	&	Transition	&	Frequency (GHz)	& V1057 Cyg &	V1515 Cyg	&	V1647 Ori	&	V1735 Cyg	&	V2492 Cyg}
\startdata
\ce{^{13}CO}	&	J=2-1	&	220.399	& \Checkmark &	\Checkmark	&	\Checkmark	&	\Checkmark	&	\Checkmark	\\
\ce{C^{18}O}	&	J=2-1	&	219.560	& \Checkmark&	\Checkmark	&	\Checkmark	&	\Checkmark	&	\Checkmark	\\
\ce{C^{17}O}	&	J=2-1	&	224.714	& \Checkmark&	\Checkmark	&	\Checkmark	&	\Checkmark	&	\Checkmark	\\
\hline
SO	&	4(5)- 3(4)	&	206.176	& \Checkmark&	\Checkmark	&	\Checkmark	&	\Checkmark	&	\Checkmark	\\
SO	&	6(5)-5(4)	&	219.949	& \Checkmark&	\Checkmark	&	\Checkmark	&	\Checkmark	&	\Checkmark	\\
\hline
\ce{SO2}	&	12(0,12)-11(1,11)	&	203.392	& \Checkmark &	\AsteriskThin	&	\XSolidBrush	&	\XSolidBrush	&	\XSolidBrush	\\
\ce{SO2}	&	11(2,10)- 11(1,11)	 &	205.301	& \Checkmark  &	\XSolidBrush	&	\XSolidBrush	&	\AsteriskThin	&	\XSolidBrush	\\
\hline
\ce{H2CO}	&	3(0,3)-2(0,2)	& 	218.222	& \Checkmark &	\Checkmark	&	\Checkmark	&	\Checkmark	&	\Checkmark	\\
\ce{H2CO}	&	3(1,2)-2(1,1)	&	225.698	& \Checkmark &	\Checkmark	&	\Checkmark	&	\Checkmark	&	\Checkmark	\\
\ce{H2CO}	&	3(2,2)-2(2,1)	&	218.476	& \Checkmark &	\XSolidBrush	&	\XSolidBrush	&	\AsteriskThin	&	\Checkmark	\\
\ce{H2CO}	&	3(2,1)-2(2,0)	&	218.760	& \Checkmark &	\XSolidBrush	&	\XSolidBrush	&	\Checkmark	&	\AsteriskThin	\\
\enddata
\tablecomments{All of the above molecules and transitions have a bright detection in source V1057 Cyg and across all other sources, either have a bright detection ($>$3$\sigma$, denoted by \Checkmark), a tentative detection ($\approx$3$\sigma$, \AsteriskThin ), or no detection (\XSolidBrush). Every other transition found in V1057 Cyg (Table \ref{tab:Bonus_lines}) is not detected in any other source.}
\end{deluxetable*}

\subsection{Line Detections}

%\begin{figure*}
%    \centering
%    \includegraphics[width=1.1\textwidth]{images/Line_rich_example.pdf}
%    \caption{\textit{An example of the spectra of the central pixel of each source between 218.2-219 GHz. This region shows how V1057 Cyg is markedly more line-rich than each of the other sources. }}
%    \label{fig:line-rich-example}
%\end{figure*}

\begin{figure*}
    \makebox[\textwidth][c]{\includegraphics[width=1.13\textwidth]{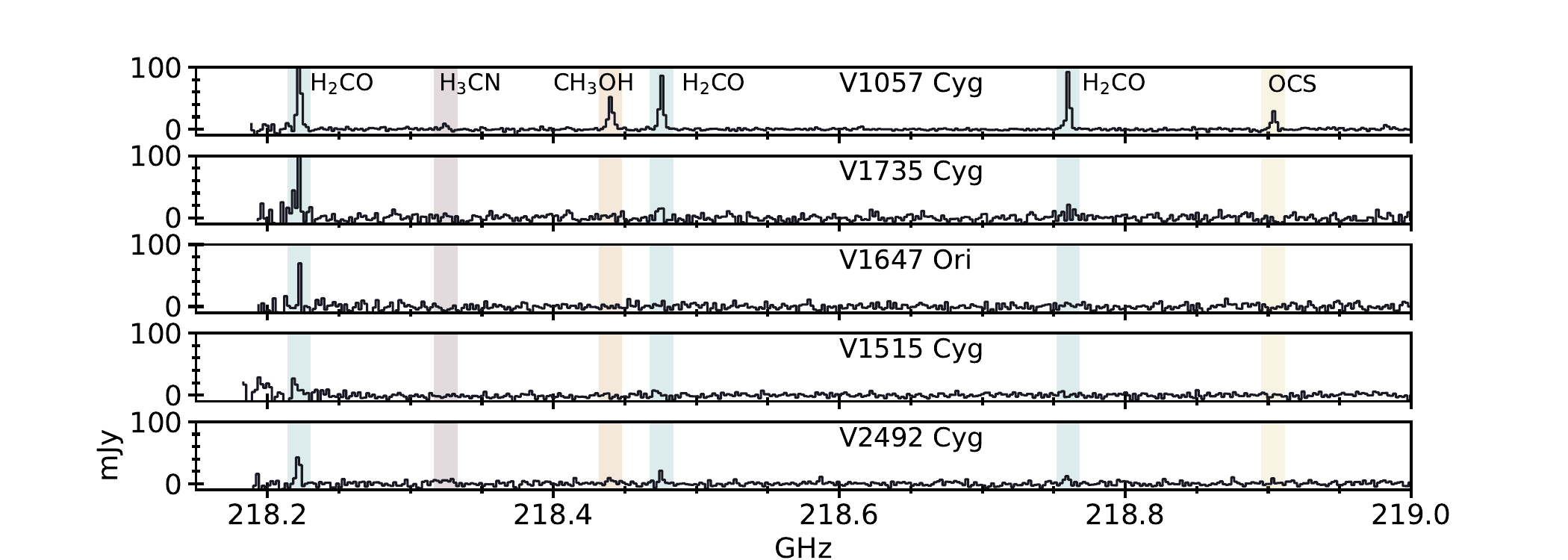}}
    \caption{\textit{An example of the spectra of the central pixel of each source between 218.2-219 GHz. This region shows how V1057 Cyg is markedly more line-rich than each of the other sources. }}
    \label{fig:line-rich-example}
\end{figure*}

\begin{figure*}
    \makebox[\textwidth][c]{\includegraphics[width=1.05\textwidth]{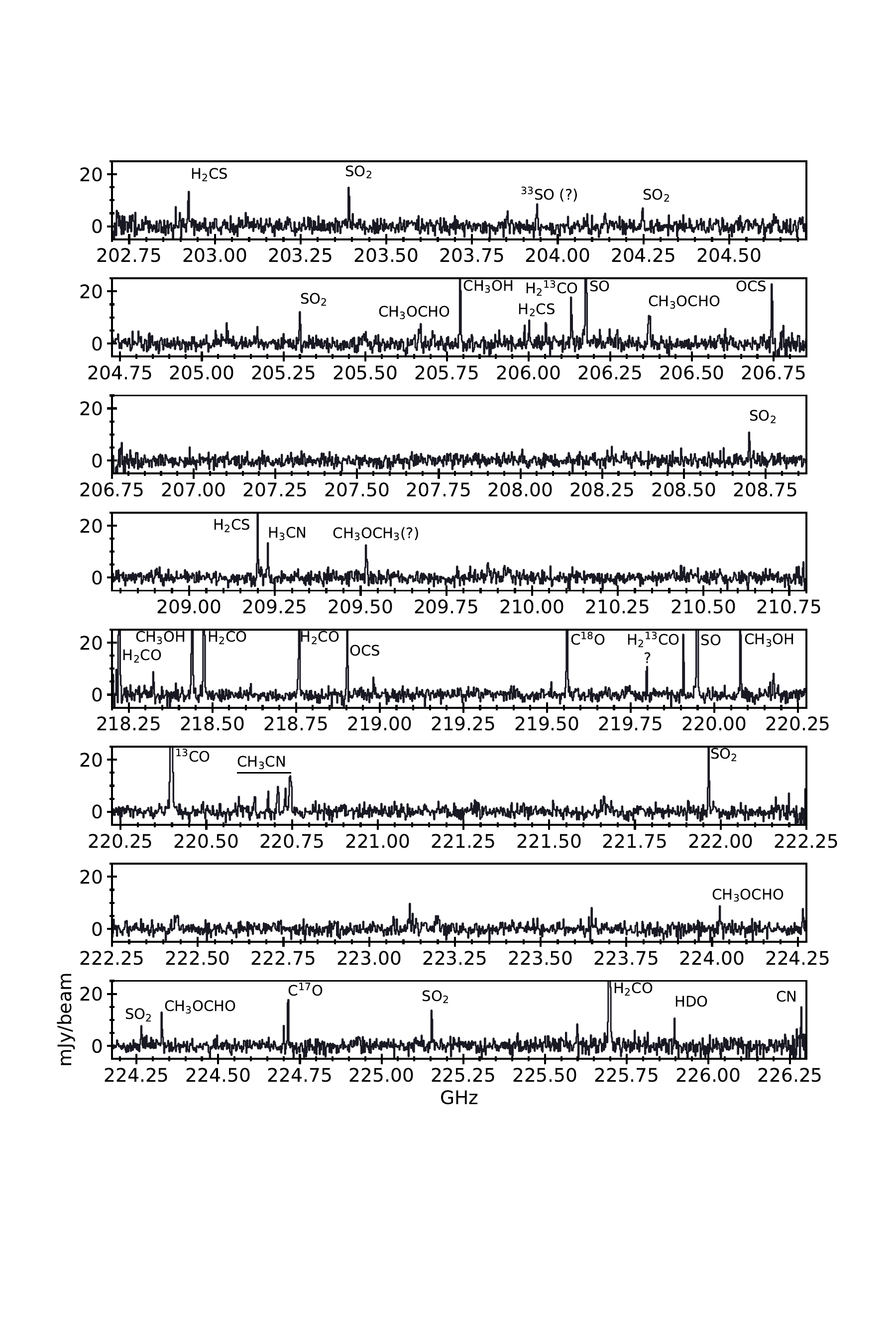}}
    
    \vspace{-5cm}
    \caption{\textit{The full lower and upper sideband coverage of V1057 Cyg. Lines that are used to calculate column densities or mass are highlighted. Lines identified with a "?" are tentative detections, mostly due to only one transition identified and it being a rare molecule to detect in protoplanetary disks. }}
    \label{fig:all_v1057cyg}
\end{figure*}

In addition to CO isotopologues, all of our sources have centrally peaked emission from SO (transitions 4$_{5}$-3$_{4}$ and 6$_{5}$-5$_{4}$) and \ce{H2CO} (transitions 3$_{03}$-2$_{02}$ and 3$_{12}$-2$_{11}$). Additional lines of \ce{H2CO} and \ce{SO2} are detected with varying degree of certainty across the sample (see Table \ref{tab:Common lines} and Figure \ref{fig:line-rich-example}). Beyond our expected science lines, serendipitous line detections were numerous over the large bandwith that NOEMA offers. %To identify bright lines, we focus on each line with a high signal-to-noise and identify a small range in which this line is emitting at given our uncertainties in spectral resolution. 
Using \textsc{splatalouge} we search for lines within this range and identify the most likely transition to be associated with the observed line. The most likely molecule will  be a molecule/transition that has already been observed, especially within the context of protoplanetary disks/FUOrs (i.e. \ce{H2CO}, \ce{CH3CN}, HDO). \ce{CH3OH} is the most common complex organic molecule (COM) when considering the ice and gas reservoir combined, and if an ice sublimation signature was left behind in the gas, \ce{CH3OH} should be present. In three out of five of our sources, we have no detection of any \ce{CH3OH} transitions, in V2492 Cyg we have two 2-3$\sigma$ detections of \ce{CH3OH} transitions. A few of our identified lines have only been detected in a few protoplanetary disks, but have been widely seen in protostars (SO, \ce{SO2}, \ce{CH3OCHO}). In this paper we only present molecular detections that have ideally multiple transitions with peak fluxes $>$4$\sigma$).

V1057 Cyg stands out in this sample as the most molecule-rich, by far (see  Figure \ref{fig:line-rich-example},\ref{fig:all_v1057cyg}). It appears to have released the molecular complexity that is expected to reside within the ice \citep[i.e.][]{McClure23}, and these sublimated ices have remained in the gas perhaps since its outburst in 1970. Here, we identify bright lines that have a high signal-to-noise (all $>3\sigma$), so that the line is clearly visible by-eye, and identify each transition that will be used to constrain an excitation temperature and column density (see Table \ref{tab:Bonus_lines} in appendix). In addition the aforementioned CO isotopologues, molecules that have robust, $>>$3$\sigma$ detections include water (HDO), complex organic molecules (COMs: \ce{CH3OH}, \ce{CH3CN}, \ce{CH3OCHO}) the building blocks for COMs (\ce{H2CO}, \ce{HC3N} ), including an isotopologue (\ce{H2}\ce{^{13}CO}), and sulfur bearing molecules (\ce{SO2}, SO, OCS, \ce{H2CS}). %This is very similar to another outbursting source: V883 Ori \citep{Lee19}. 

\subsection{Column Density \& Temperature Extraction from CASSIS}

\colorlet{color1}{green!10!white}
\colorlet{color2}{cyan!10!white}
\colorlet{color3}{red!10!white}
\colorlet{color4}{olive!10!white}

\begin{deluxetable}{ccc}
\label{tab:col_dens}
\tablecolumns{3}
\tablewidth{0pt}
\tabletypesize{\small}
\tablecaption{Column Densities for Detected Molecular Species in V1057 Cyg}
\tablehead{Molecule		&	T$_{ex}$ & Column Density \\
		&	[K] & [  mol/cm$^{2}$] }
\startdata
\rowcolor{color3} \multicolumn{3}{c}{Sulfur Species}\\					
\ce{SO2}	&	80	&	1.6($\pm$ 0.32)$\times 10^{14}$	\\
\ce{H2CS}	&	80	&	4.0($\pm$ 0.80)$\times 10^{13}$	\\
\ce{SO}	&	120$^{a}$	&	2.5($\pm$ 0.50)$\times 10^{14}$	\\
\ce{OCS}	&	80$^{a}$	&	1.6($\pm$ 0.32)$\times 10^{14}$	\\
\rowcolor{color4} \multicolumn{3}{c}{Building-block COMs}\\					
\ce{CH3CN}	&	110	&	1.0($\pm$ 0.20)$\times 10^{13}$	\\
\ce{H2CO}	&	90	&	2.5($\pm$ 0.50)$\times 10^{14}$	\\
\ce{CH3OH}	&	180	&	3.2($\pm$ 0.64)$\times 10^{15}$	\\
\ce{H3CN}	&	120$^{a}$	&	2.5($\pm$ 0.50)$\times 10^{14}$	\\
\ce{H2^{13}CO}	&	80$^{a}$	&	1.6($\pm$ 0.32)$\times 10^{13}$	\\
\ce{H2C^{17}O}	&	80$^{a}$	&	1.6($\pm$ 0.32)$\times 10^{13}$	\\
\rowcolor{LightYellow} \multicolumn{3}{c}{Long COMs}\\					
\ce{CH3OCHO}	&	120	&	2.5(0.5)$\times 10^{14}$	\\
\rowcolor{color2} \multicolumn{3}{c}{Water}\\					
HDO	&	N/A	&	2.0(0.40)$\times 10^{14}$	\\
\enddata					
\tablecomments{ Column density and temperature with the lowest $\chi^{2}$ are reported. 2$\sigma$ errors derived from the $\chi^{2}$ minimization given a fixed temperature gave column density uncertainties less than 20\%, which is believed to be less than the calibration uncertainty, thus 20\% errors are stated.  $^{a}$These molecules had 1-2 strong transitions and the temperatures 80, 120, and 180~K were used to determine a column density. Out of these models, the one with the lowest $\chi^{2}$ is reported in this table.}
\end{deluxetable}

We use CASSIS\footnote{http://cassis/irap.omp.eu} \citep{Vastel15} which is a slab modeling tool for spectral analysis to determine column densities and excitation temperatures for the molecules detected in V1057 Cyg. We assumed LTE and utilized the line lists form the JPL line catalogue\footnote{https://spec.jpl.nasa.gov/home.html}. For each of the molecules with three or more transitions, we perform a grid exploration of column density, temperature, and full-width half-max (FWHM) of the line. For each line we explore column densities between 10$^{13}$ and 10$^{18}$ cm$^{-2}$, excitation temperatures between 40 and 300~K, and FWHM between 3 and 7 km/s. We include lines in our slab models for any transition that has an A$_{ij}$ greater than 10$^{-6}$ and all transitions with an E$_{up}$ lower than 500~K, so that both line detections and non-detections can constrain our final column density and temperature. A $\chi^{2}$ fitting routine was done to determine the slab model within the explored parameter space that fit the data best. Both line detections and non-detections were used to constrain a column density and temperature for each molecule that had three or more transitions. Temperature is not well constrained, between the explored range solutions within a 1$\sigma$ certainty exist, given a fixed temperature however the column densities can be constrained. To calculate our errors, we fix our slab model to the 40 and 300~K and determine the lowest and highest column density within a 1$\sigma$ certainty, respectively.  In V1057 Cyg, molecules with three or more transitions were \ce{CH3OH}, \ce{H2CO}, \ce{SO2}, \ce{CH3CN}, \ce{H2CS}, and \ce{CH3OCHO}. The final column densities and temperatures from the slab model with the smallest $\chi^{2}$ are shown in Table \ref{tab:col_dens}.

%Given the posterior distribution of the $\chi^{2}$ exploration, the temperature is not well constrained, with 1$\sigma$ values extending to the maximum and minimum temperature values that could be allowed (40-300~K). The column density is more narrowly constrained. Considering the full possible temperature range, a column density up to a factor of two above/below our reported column densities provide fits to the data within 2$\sigma$ certainties.  

%The two sulfur bearing species with 3+ transitions (\ce{H2CS} and \ce{SO2}) seem to originate from a nearly identical temperature (80~K), while the small COMs come from a slightly higher temperature (90-180~K). 

The rest of our detected molecules have 1-2 transitions, which would not provide enough strong constraints on a parameter exploration to determine their column densities and excitation temperatures. Thus, we choose to keep the temperature fixed, and iterate over the same column density range as before. We do this $\chi^{2}$ minimization over three different fixed temperatures (80~K, 120~K, 160~K), motivated by the temperatures found for molecules with 3+ transitions.  For each temperature and column density combination we calculate the  $\chi^{2}$, and report back the temperature and column density that has the lowest overall $\chi^{2}$ value. 
%OCS appears to emit from a similar temperature as SO$_{2}$ and H$_{2}$CS while SO may come from a warmer region. The formaldehyde isotopologues appear to emit from the same temperature as the main H$_{2}$CO emission. HC$_{3}$N originates from warm temperatures, akin to \ce{CH3CN}. 
One line of HDO is not enough to put a constraint on temperature, but across 80, 120, and 180~K a column density of 1.0$\times$10$^{14}$ (cm$^{-2}$) returned the lowest $\chi^{2}$. The fact that these molecules are detected at relatively high abundances, and at temperatures above their sublimation temperature points to these volatiles being sublimated off of grains.%.All of these results are consistent with these volatiles being sublimated off of the grains and into the gas phase. 

We also calculate upper limits on the \ce{CH3OH} column density for each source that has non-detections of any \ce{CH3OH} lines. LTE slab models are created using the derived excitation temperature from V1057 Cyg (180~K). A series of column densities are used, and when a slab model with a certain column density creates a line that should have been detected at the 3$\sigma$ level, we use that as our upper limit for the \ce{CH3OH}. To provide context to the derived methanol column density, we also derive C$^{17}$O and \ce{H2CO} column densities. \ce{H2CO} has two or more strong detections in each of our sources, thus we use a $\chi^{2}$ fitting technique.   More on calculating the column density, and  the optically thin emission from C$^{17}$O as a mass tracer can be found in \citet{Calahan24}.

\begin{deluxetable*}{cccccc}
\label{tab:upper_limts}
\tablecolumns{3}
\tablewidth{0pt}
\tabletypesize{\small}
\tablecaption{Column Densities and Mass Across Sample}
\tablehead{Molecule		&	V1057 Cyg & V2492 Cyg & V1735 Cyg & V1515 Cyg & V1647 Ori}
\startdata
%\rowcolor{color3} \multicolumn{3}{c}{Sulfur Species}\\					
\ce{CH3OH}	& 3.2$\times 10^{15}$	& 1.0$\times 10^{15}$ & $<$1.5$\times 10^{15}$ &  $<$1.0$\times 10^{15}$ &  $<$5.0$\times 10^{14}$	\\
\ce{H2CO}	& 2.5$\times 10^{14}$	& 2.5$\times 10^{14}$ & 6.30$\times 10^{13}$	& $\sim$6.3$\times 10^{13}$ & $\sim$2.5$\times 10^{13}$\\
\ce{C^{17}O}	& 2.0$\times 10^{17}$	& 7.0$\times 10^{17}$ & 1.9$\times 10^{16}$	& 5.5$\times 10^{16}$&  1.5$\times 10^{17}$\\	
Inner Region Mass [M$_{\odot}$] & 3.4 & 1.9 & 0.25 & 0.58 & 0.42\\
\enddata
\tablecomments{Derived column densities or upper limits for each source (mol/cm$^{2}$); see Figure \ref{fig:CH3OH_compare}. The derivation for inner region mass including envelope and disk emission can be found in \citet{Calahan24} %of \ce{CH3OH} for V1057 Cyg and V2492 Cyg, and upper limits for V1515 Cyg, V1735 Cyg, and V1647 Ori. \ce{H2CO} column densities for V1647 Ori and V1515 Cyg were determined using only two transitions. 
}
\end{deluxetable*}

\begin{figure}
    \centering
    \includegraphics[width=.49\textwidth]{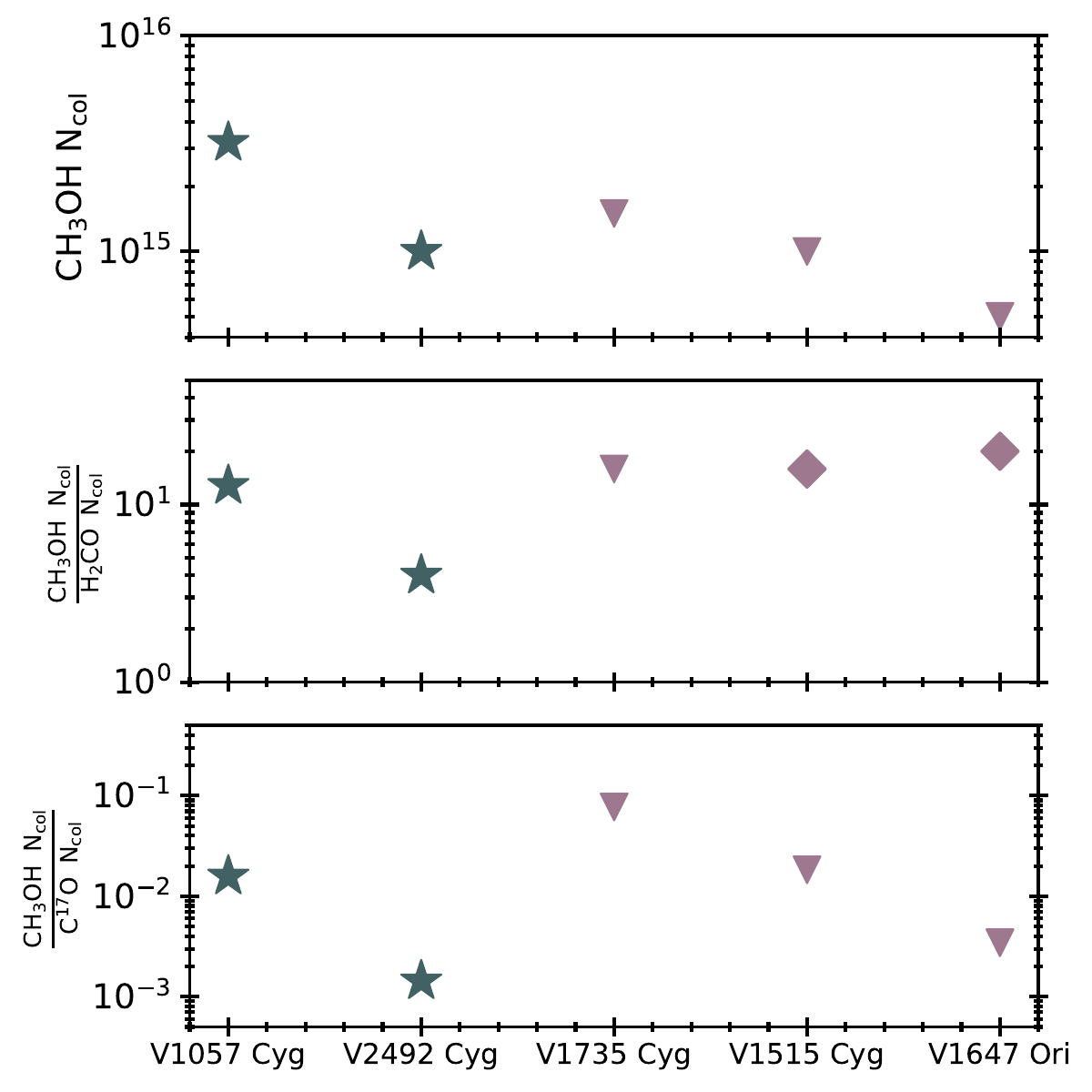}
    \caption{\textit{Column densities of \ce{CH3OH} in V1057 Cyg (based on 5 transitions) and V2472 Cyg (based on two transitions) and derived upper limits from the non-detections towards the three other sources (top). The ratio between the \ce{CH3OH} column density or upper limit and a derived \ce{H2CO} column density (middle) and C$^{17}$O column density (bottom). Points with stars indicate that detected line(s) are used to determine the column density. A triangle indicates an upper limit, and diamond utilized both an upper limit on the methanol and a \ce{H2CO} column density based on just two transitions.}}
    \label{fig:CH3OH_compare}
\end{figure}

\section{Analysis \& Discussion}\label{sec:diss}

\subsection{Line Accounting}

Beyond the CO lines, at least two bright formaldehyde transitions are detected ubiquitously across the sample. \ce{H2CO} detections are not particularly surprising, as this is a common molecule found within both protoplanetary disks and earlier Class 0/I sources \citep[i.e.][]{Pegues20,Jorgensen07}. \ce{H2CO} emission is associated with warm gas and shocks, and seems to be present in disk environments over a wide range of gas masses. Ubiquitous detections of SO and tentative detections of SO$_{2}$ in three of our sources however, is surprising.  SO and \ce{SO2} are have been found in only four quiescent Class II protoplanetary disks - all massive disks around Herbig Ae/Be stars \citep{Riviere-Marichalar20,Booth21b,Law23,Booth23}. Sulfur-bearing molecules are found more often within protostars, however the explanation for its origin can come from shocks or infall, releasing these molecules from ice grains \citep[i.e.][]{ArthurdelaVillarmois19,LeGal20}. In our outburst sources, SO (and \ce{SO2}) detections are bright and centrally peaked with only a mild asymmetry in the line wings. This suggests that the majority of the sulfur was released from ices within the disk for each of our sources, as opposed to fast moving shocked gas at a very different velocity than the source. The presence of SO can come from the outburst event, leaving a sublimated ice-signature behind or it has also been shown that SO and \ce{H2CO} can be enhanced in gravitationally unstable disks, due to heat from spiral structure and shocks \citep{Evans15, Evans19}. 

%In this sub-section we explore wheather this implies a different physics or chemistry between V1057 Cyg and the other sources, or can be attributed to there being less overall material around the four line-poor sources.
%those listed in Table \ref{tab:Common lines}
%Only a tentative 2$\sigma$ detection of a single \ce{CH3OH} line can be found in V2492 Cyg. 
%An increased abundance of gas-phase \ce{CH3OH} in FU Ori type outbursting sources would be expected, as this is the most common COM in the disk as a whole, but typically is frozen out onto the ice for the vast majority of the spatial extent of a disk around a quiescent star. In an outbursting source where the temperature is high and remains high as the star returns to its quiescent luminosity, \ce{CH3OH} should be abundant in the gas-phase for a much larger extent of the disk, thus more easily observable.  Thus, it was unexpected that only one of our outbursting sources have clear detections of \ce{CH3OH}. 
%V1057 Cyg does contain the most gas mass out of our sources within $\sim$1000~au of the star \citep{Calahan24}. V2492 Cyg is the second most massive source in our sample, and it has a tentative detection of \ce{CH3OH} with two weak transitions. Our other sources have no detections of \ce{CH3OH}. 

Four of our sources (V1515 Cyg, V1735 Cyg, V1647 Ori, and V2492 Cyg) are void of bright lines outside of CO isotopologues, \ce{H2CO}, SO, and tentatively \ce{SO2}. Notably, none of them exhibit strong lines of the most common COM - \ce{CH3OH}.
In order to determine if the reason behind the \ce{CH3OH} detection in V1057 Cyg comes from its high mass, or something unique about this system we determine upper limits on the \ce{CH3OH} column density for our other sources. The upper limits on \ce{CH3OH} are shown in Figure \ref{fig:CH3OH_compare}. For V2492 Cyg, we have two 2-3$\sigma$ detected transitions of \ce{CH3OH} thus we run slab models to determine if 80, 120, or 180~K provide the best fit to the data. We find that we get the best model with an excitation temperature of 120~K and a column of 1$\times$10$^{15}$ cm$^{-2}$, about a factor of three less than what was found in V1057 Cyg. All other methanol column density upper limits are also determined to be lower than V1057 Cyg. 

\subsection{An Apparent Dichotomy in Chemical Signatures}
When comparing spectral data from V1057 Cyg to our other sources, V1057 Cyg appears to have a markedly different chemistry. The chemical richness is representative of what we expect from ice sublimation shortly after an outburst. The fact that it is the only source in our sample that appear to have this chemical signature was not expected based on the $\sim$two outbursting sources that have been observed previously, V883 Ori and L1551 IRS5. The fact that four of our sources are not line-rich brings up the question as to whether there is something unique about the outburst or disk environment in V1057 Cyg that keeps this chemical signature around for 50+ years, while other sources may lose this expected chemical signatures more rapidly. %However, V1057 Cyg is a unique source in our sample in several ways which must be considered in context with the observed chemical signature. 

From C$^{17}$O observations, V1057 Cyg is the most massive in our sample, and the mass derived from optically thin C$^{17}$O is markedly higher than what would be derived from continuum emission. This, in addition to a lack of clear disk features on the $\sim$1000s of au scales in position-velocity space, suggests that V1057 Cyg has a significant envelope contribution, implying that V1057 Cyg may be closer to a Class I object (although previous work has identified it as a Class II source based on the SED). Adding to this source's complexity, it is also a binary system, as seen from two peaks in the continuum observations \citep{Calahan24} and as was first spatially resolved by \citet{Green16}. The molecular emission from all non-CO lines appear as unresolved emission, peaked on the central star. Thus, we expect these molecular detections to be solely coming from the central protostar's disk/envelope.

When comparing the \ce{CH3OH} column density in V1057 Cyg to our other four sources, it is 2-3 times larger than our upper limits. To provide context to the \ce{CH3OH} column, we take the ratio between \ce{CH3OH} and the one other organic that is found in all of our sources, \ce{H2CO} as well as the optically thin gas mass tracer \CsvO{} (see Fig. \ref{fig:CH3OH_compare}). Now, we see that the \ce{CH3OH} in the other sources are comparable to V1057 Cyg or below. %Given our upper limits, V1735 Cyg, V1515 Cyg, and V1647 Ori all have higher limits on the column density ratios between \ce{CH3OH} and \ce{H2CO} than what is found for V1057 Cyg. 
The limits on the other sources show that there is not a significantly higher ratio of \ce{CH3OH}/\ce{H2CO}. Compared to the one other source with tentative \ce{CH3OH} detections, V2492 Cyg, V1057 Cyg does appear to have significantly more \ce{CH3OH}, even within the context provided by \ce{H2CO}. Deeper observations of bright \ce{CH3OH} lines or in a frequency range with optically thinner dust would be required to determine if there is a clear dichotomy in the chemical signatures across FU Ori sources. Considering \ce{CH3OH}/\CsvO{}, it does appear that the upper limit found in V1647 Ori is significantly lower than in V1057 Cyg, while the other two sources place the upper limit of \ce{CH3OH}/\CsvO{} on order of V1057 or higher. This data set alone cannot definitively prove that V1057 Cyg has a unique chemical signatures amongst outbursting sources,  its apparent molecular complexity may be due to its high gas mass within the surrounding $\sim$1000~au. It is peculiar that the source with the third highest gas mass and comparable dust mass (V1647 Ori) has no \ce{CH3OH} and only two detections of \ce{H2CO}. To say anything definitive about the gaseous inventory of V1735 Cyg and V1515 Cyg will require deeper observations.  

\subsection{Complex Organic Content Evolution: Protostars to Comets}

\begin{figure*}
    \centering
    \includegraphics[width=.8\textwidth]{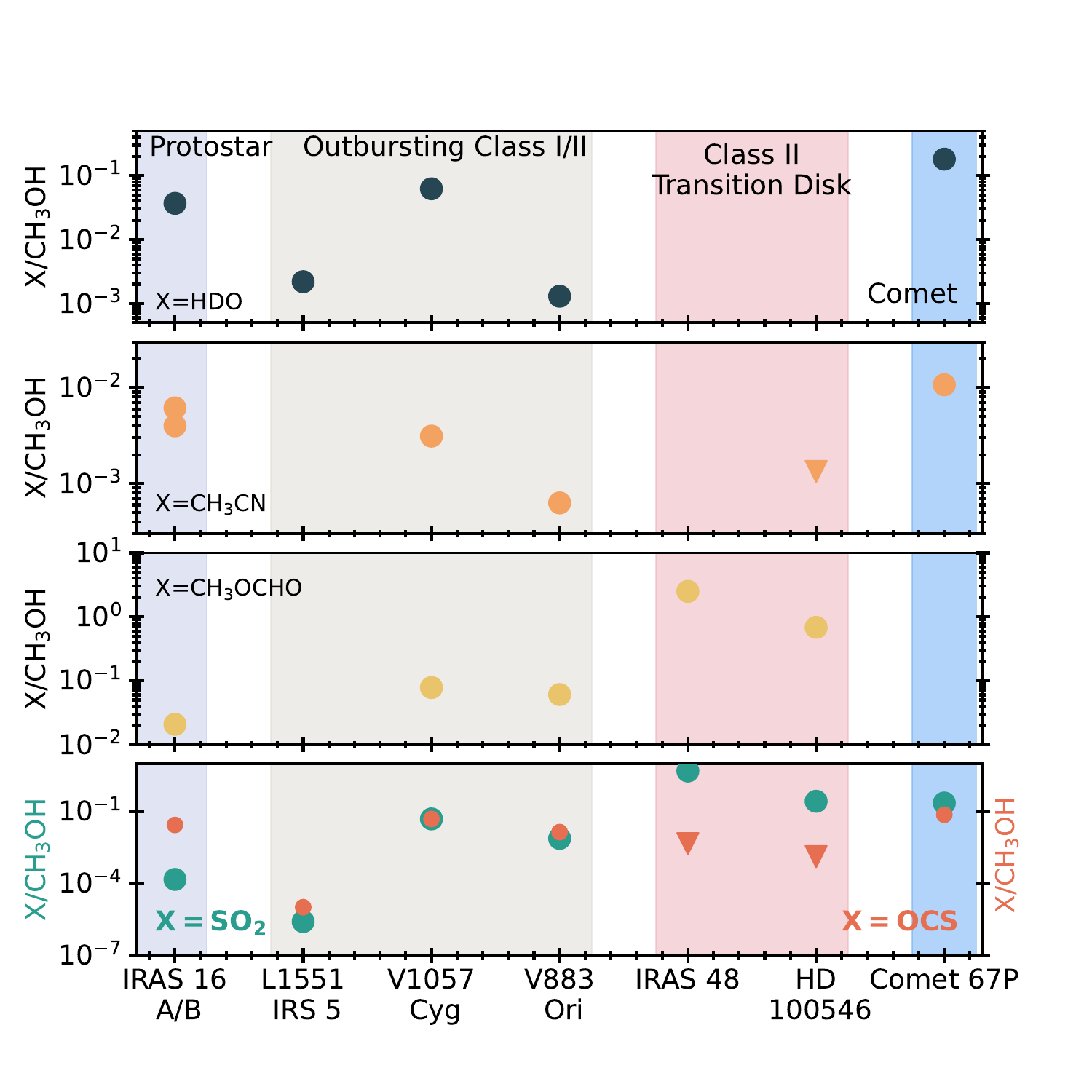}
    \caption{Column density ratios across protostars, outbursting sources, a protoplanetary disk, and abundance ratio from a comet of molecules with respect to CH$_{3}$OH. IRAS 16293-2422 is a binary system and both A and B are plotted along the same x-value.}
    \label{fig:compare_across_evol}
\end{figure*}

Chemical inheritance from cloud to comet is essential to constrain the critical factors that influence final planetary compositions. The chemical inventory within V1057 Cyg is rich enough to compare to protostars, other outbursting sources, protoplanetary disk, and comet data across multiple molecules (see Figure \ref{fig:compare_across_evol}). 

The PILS Survey is an example of the most well studied protostellar system, as it targeted the prostostar IRAS 16293-2422 across a wide spectral range. This system is a binary, and both protostars appear to be quite line-rich with HDO, complex organic molecules, and sulfur-bearing species. Compared to V1057 Cyg alone, the molecular ratio between each of these molecules and \ce{CH3OH} is lower than V1057 Cyg, with the exception of \ce{CH3CN} in IRAS 16293-2422A \citep{ Persson13,Calcutt18, Jorgensen18, Drozdovskaya18,Manigand20}. 

There are two other outbursting sources that have been classified as an FU Ori type object or FUOr-like and have been seen to be line-rich, which are V883 Ori and L1551 IRS 5 respectively. Based on the relative amount of envelope in each source and historical SED classifications, L1551 IRS 5 likely is the youngest out of these three, followed by V1057 Cyg, then V883 Ori. It is worth noting that between the sources that have some long-term lightcurve data, they have different trends over time. V883 Ori likely went into outburst before 1888 based on photometric plate data\citep{Pickering1890}, and since its discovery has remained very luminous \citep{Dent98, Greene08}. V1057 Cyg's outburst, however was captured as a sharp and quick rise, while then decreasingly fairly quickly at first, and then slowly yet steadily till present day\citep{Szabo21}. Both of these sources have existed post-outburst for $\sim$100-200 years, which is much shorter than a typical freeze-out time \citep[1000+ years][]{Visser15}. Thus even though these sources have markedly different histories, the fact that they have been post outburst within the past $\sim$100 years leads us to assume that the chemistry between the sources has not had time to be altered significantly. HDO appears to be by far the most abundant in V1057 Cyg, with the ratio of HDO/\ce{CH3OH} nearly two orders of magnitude higher than in L1551 IRS 5 and V883 Ori. For the sulfur-bearing species, V1057 Cyg and V883 Ori are on par with each other, and both the OCS and SO$_{2}$ to \ce{CH3OH} ratios are similar. L1551 IRS 5 has a much smaller relative abundance of sulfur-bearing molecules. The COMs \ce{CH3CN} and \ce{CH3OCHO} are detected in V1057 Cyg and V883 Ori, with \ce{CH3OCHO}/\ce{CH3OH} being very similar and \ce{CH3CN}/\ce{CH3OH} just under an order of magnitude \citep{Tobin23, Andreu23,Yamato24, Marchand24}. The other FUOr object that has been targeted with NOEMA, HBC494, has a measured ratio between \ce{CH3CN} and \ce{CH3OH}, which is a factor of ten higher than what is found in V1057 Cyg.

Gas-rich protoplanetary disks around T Tauri stars will have a very small emitting area and gas colum in which sublimated COMs, HDO, and sulfur-bearing molecules can be observed. Transitions disks around Herbig Ae/Be stars, however, offer a larger emitting area due to both the inner radius location and higher temperature. With the addition of typically higher gas masses, these objects offer the best insight into the chemistry available during the Class II stage of a protoplanetary disk. IRS 48 and HD 100546 are two transition disks which have been well studied and column densities of these sublimated molecules can be studied. In both HD 100456 and IRS 48, there are no detected \ce{CH3CN} and observations were not sensitive to HDO. Compared to the protostar and FU Ori sources, the protoplanetary disks have much higher relative abundances of the long COM \ce{CH3OCHO} and SO$_{2}$, while the upper limit on OCS is on par with the slightly more evolved FUOrs \citep{Booth24, Booth24b}.  

Finally, Comet 67P/Churyumov-Gerasimenko has direct detections of water, \ce{CH3CN}, OCS, and SO$_{2}$. The sulfur bearing species are on par with the FUOrs, water is slightly higher than V1057 Cyg, but signifcantly higher than the other FUors and protostar. \ce{CH3CN} in the comet is most abundant with respect to \ce{CH3OH} as compared to all other sources in this comparison \citep{LeRoy15, Altwegg17, Rubin19}. 

V1057 Cyg adds a critical data point between protostars, protoplanetary disks and comets. Given the limited data points, there appears to be a possible trend in the sulfur-bearing molecules and \ce{CH3OCHO}, as they increase in relative abundance across stellar age. If true, this would suggest that large COMs and simple sulfur bearing molecules build up over time from protostar to comet. HDO and \ce{CH3CN} show no clear trend. HDO especially shows a wide range in relative abundance to \ce{CH3OH}, spanning across two orders of magnitude within the outbursting Class I/II sources alone. This $>$ order of magnitude range in the HDO/\ce{CH3OH} ratio is striking, and could be from optically thick \ce{CH3OH}, an enhanced water or depleted \ce{CH3OH} reservoir, or an enhanced D/H in water or a combination of these. Observations of isotopes of \ce{CH3OH} and of H$_{2}^{18}$O would shed light into which of these is most likely.  More data points and wider spectral surveys of multiple line-rich objects are necessary to understand if these relative abundance trends hold, and how they may impact the chemistry available to forming planets over time.

\section{Conclusion}\label{sec:conclusion}

Using NOEMA, we targeted 5 FUOr and EX Lup type sources between 202.7-210.8 GHz and 218.2-226.3 GHz. All of our sources contained emission from $^{13}$CO, C$^{18}$O, C$^{17}$O, SO, and H$_{2}$CO. One of our sources, V1057 Cyg, has a markedly different spectra, rich with volatiles that one would expect to see post-outburst when the ice has sublimated. This is akin to another FUOr, V883 Ori. This survey more than doubles the number of FUOr outburst objects that have been surveyed across a wide spectral range, yet less than half exhibit the assumed chemical signature of an outburst: V1057 Cyg, V883 Ori, and L1551. %Given these observations, we are unable to definitively conclude that the other four sources in our sample have a markedly different chemistry than V1057 Cyg and V883 Ori, however upper limits suggest the chemical inventory is on par or less rich in the gas phase than these brighter more massive sources (V1057 Cyg and V884 Ori).
We explored whether these observed differences could be explained by differences in gas mass within the beam, and find that they could. Deeper observations are needed to investigate the range of ice compositions present in planet forming disks around young stars. 
In these four other sources in our sample, the molecular richness from sublimated ices could have frozen back onto grains quickly, or our observations did not go deep enough to obtain the sensitivity required.%, or the volatile-rich gas may be hidden behind optically thick dust in this frequency range. 
V1057 Cyg, V883 Ori, and L1551  have envelope emission, signifying that these are younger sources than other FU Or/Ex Lup objects and this could be a factor in determining the abundance of volatile-rich gas. Both sources have peak luminosities of $>200$L$_{\odot}$ suggesting a larger area of sublimated species, however classic FUOrs V1515 Cyg and V1735 Cyg in our sample have comparably bright peak luminosities (see Table \ref{tab:SourceProperties}). More observations of outbursting sources will offer a unique glimpse into the chemistry available to forming planets, and deeper observations would greatly benefit our understanding of chemical inheritance. 

% A molecular inventory of each source was taken, and it was found that V1057 Cyg was uniquely molecularly rich, with long COMs, four different sulfur-bearing molecules, and a water detection. The other four sources only had \ce{H2CO}, \ce{SO}, and tentative \ce{SO2} detections in addition to the CO. This could suggest either a reset-chemistry in certain out-bursting environments or a quick re-freeze-out of volitle molecules. More follow-up observations and modeling should be done to explain this pattern. 

%\acknowledgments
\section{Acknowledgments}

Matplotlib \citep{Matplotlib}, Astropy \citep{astropy:2013,astropy:2018}, NumPy \citep{harris2020array}

J.K.C. acknowledges support from the National Science Foundation Graduate Research Fellowship under Grant No. DGE 1256260 and the National Aeronautics and Space Administration FINESST grant, under Grant no. 80NSSC19K1534. 

A. S. B. is supported by a Clay Postdoctoral Fellowship from the Smithsonian Astrophysical Observatory.

E.A.B. acknowledge support from NSF Grant\#1907653 and NASA grant XRP 80NSSC20K0259

%\end{linenumbers}

\appendix
\restartappendixnumbering
\renewcommand{\thetable}{A\arabic{table}}

\begin{deluxetable*}{cccccc}
\label{tab:Bonus_lines}
\tablecolumns{3}
\tablewidth{0pt}
\tabletypesize{\small}
\tablecaption{Molecular Transitions In V1057 Cyg Used for Slab Models Excluding CO Isotopologues}
\tablehead{Molecule	&	Transition	&	Frequency (GHz) &	log$_{10}$A$_{ij}$	&	E$_{\rm up}$ (K)	&	V1057 Cyg Detection?}
\startdata
\ce{CH3OH}	&	12(5)-13(4)	&	206.001	&	-4.98	&	317	&	yes	\\
	&	1(1,1)-2(0,2)++	&	205.791	&	-4.473	&	16.8	&	yes	\\
	&	19(1)-19(0)	&	209.518	&	-4.500	&	462	&	yes	\\
	&	4(2,2)-3(1,2)	&	218.44	&	-4.329	&	45.5	&	yes 	\\
	&	8(0,8)-7(1,6)	&	220.078	&	-4.599	&	96.6	&	yes	\\
	&	10(5,6)-11(4,8)	&	220.401	&	-4.950	&	252	&	yes	\\
\ce{H2CO}	&	3(0,3)-2(0,2)	&	218.222	&	-3.550	&	21.0	&	yes	\\
	&	3(2,2)-2(2,1) 	&	218.475	&	-3.595	&	68.1	&	yes	\\
	&	3(2,1)-2(2,0)	&	218.76	&	-3.802	&	68.1	&	yes	\\
	&	3(1,2)-2(1,1)	&	225.697	&	-3.557	&	33.5	&	yes	\\
\ce{CH3CN}	&	12(0)-11(0)	&	220.747	&	-2.180	&	68.9	&	yes	\\
	&	12(1)-11(1)	&	220.743	&	-2.193	&	76.0	&	yes	\\
	&	12(2)-11(2)	&	220.73	&	-2.234	&	97.4	&	yes	\\
	&	12(3)-11(3)	&	220.709	&	-2.000	&	133	&	yes	\\
	&	12(4)-11(4)	&	220.679	&	-2.397	&	183	&	yes	\\
	&	12(5)-11(5)	&	220.641	&	-2.522	&	247	&	yes	\\
	&	12(6)-11(6)	&	220.594	&	-2.377	&	326	&	tentative	\\
\ce{SO2}	&	12(0,12)-11(1,11)	&	203.391	&	-4.055	&	70.0	&	yes	\\
	&	18(3,15)-18(2,16)	&	204.246	&	-4.033	&	181	&	yes	\\
	&	7(4,4)-8(3,5)	&	204.384	&	-4.953	&	65.0	&	tentative	\\
	&	11(2,10)-11(1,11)	&	205.301	&	-4.270	&	70.2	&	yes	\\
	&	3(2,2)-2(1,1)	&	208.700	&	-4.173	&	15.3	&	yes	\\
	&	12(5,7)-13(4,10)	&	209.936	&	-4.797	&	133	&	tentative	\\
	&	22(7,15)-23(6,18)	&	219.276	&	-4.671	&	352	&	no	\\
	&	11(1,11)-10(0,10)	&	221.965	&	-3.943	&	60.4	&	yes	\\
	&	6(4,2)-7(3,5)	&	223.883	&	-4.935	&	58.6	&	no	\\
	&	20(2,18)-19(3,17)	&	224.264	&	-4.400	&	207	&	yes	\\
	&	13(2,12)-13(1,13)	&	225.154	&	-3.283	&	57.1	&	yes	\\
\ce{H2CS}	&	6(1,6)-5(1,5)	&	202.923	&	-3.926	&	47.3	&	yes	\\
	&	6(5,1)-5(5,0)	&	205.942	&	-4.410	&	361	&	no	\\
	&	6( 0, 6)- 5( 0, 5)	&	205.987	&	-3.894	&	34.6	&	yes	\\
	&	6( 4, 3)- 5( 4, 2)	&	206.001	&	-4.150	&	245	&	no	\\
	& 6( 4, 2)- 5( 4, 1)	&	206.001	&	-4.150	&	245	&	no	\\
	&	6( 3, 4)- 5( 3, 3)	&	206.051	&	-4.019	&	153	&	yes, blended	\\
	&	6( 3, 3)- 5( 3, 2)	&	206.052	&	-4.019	&	153	&	yes, blended	\\
	&	6( 2, 5)- 5( 2, 4)	&	206.053	&	-3.945	&	87.3	&	yes, blended	\\
	&	6( 2, 4)- 5( 2, 3	&	206.158	&	-3.945	&	87.3	&	no	\\
	&	6( 1, 5)- 5( 1, 4)	&	209.200	&	-3.886	&	48.3	&	yes	\\
\rowcolor{color1} \multicolumn{6}{c}{Molecules with 1-2 Transitions}\\												
\ce{SO}	&	4(5)- 3(4)	&	206.176	&	-3.987	&	38.6	&	yes	\\
	&	6(5)-5(4)	&	219.949	&	-3.865	&	35.0	&	yes	\\
\ce{OCS}	&	17-16	&	206.745	&	-4.592	&	89.3	&	yes	\\
	&	18-17	&	218.903	&	-4.517	&	99.81	&	yes	\\
\ce{H3CN}	&	23-22	&	209.23	&	-3.14	&	121	&	yes	\\
	&	24-23	&	218.325	&	-3.084	&	130.982	&	yes	\\
\ce{H2}\ce{^{13}CO}	&	3( 1, 3)- 2( 1, 2)	&	206.132	&	-3.675	&	31.6	&	yes	\\
	&	3(1,2)-2(1,1)	&	219.908	&	-3.591	&	32.9	&	yes	\\
\ce{HDO}	&	3(1,2)-2(2,1)	&	225.896	&	-4.879	&	168	&	yes	\\
\enddata
%\tablecomments{}
\end{deluxetable*}

%\begin{figure*}
%    \centering
%    \includegraphics[width=\textwidth]{images/fluxes_cols_example.pdf}
%    \caption{\textit{The integrated flux, column density of J=2 level, and total \ce{C^{17}O} for V1057 Cyg. The black contour indicates an identified ``disk'' region and the white contour indicates ``cloud''}}
%    \label{fig:fluxes_cols}
%\end{figure*}

%\begin{figure*}
%    \centering
%    \includegraphics[width=\textwidth]{images/C17O_linewidth_example.pdf}
%    \caption{\textit{The spectra of the central pixel isolating the \ce{C^{17}O} J=2-1 emission line towards V1057 Cyg. The black horizontal lines show the calculated $\pm$ RMS and the red lines and highlighted magenta line show the frequency range in which the integrated flux is calculated for the final column density calculation. }}
%    \label{fig:C17O_example}
%\end{figure*}

\bibliography{references}{}
\bibliographystyle{aasjournal}

\newpage
%\bibliography{sample63}{}
%\bibliographystyle{aasjournal}

\end{document}